\documentclass[10pt,letterpaper]{article}
\usepackage[top=0.85in,left=1.6in,footskip=0.75in]{geometry}

\usepackage{amsmath,amssymb}

\usepackage{changepage}

\usepackage[utf8x]{inputenc}

\usepackage{textcomp,marvosym}

\usepackage{cite}

\usepackage{nameref,hyperref}

\usepackage[right]{lineno}

\usepackage{microtype}
\DisableLigatures[f]{encoding = *, family = * }

\usepackage[table]{xcolor}

\usepackage{array}

\usepackage{soul}

\usepackage{caption,tabularx,booktabs}
\usepackage{pdfpages}

\newcolumntype{+}{!{\vrule width 2pt}}

\newlength\savedwidth

\usepackage[aboveskip=1pt,labelfont=bf,font=small,labelsep=period,singlelinecheck=off]{caption}

\makeatletter
\renewcommand{\@biblabel}[1]{\quad#1.}
\makeatother

\usepackage{lastpage,fancyhdr,graphicx}
\usepackage{epstopdf}
\usepackage{dblfloatfix}
\usepackage{xcolor} 
\usepackage{subcaption}	
\usepackage{multicol}
\usepackage{graphicx}	
\usepackage{mwe}
\usepackage{textcomp}
 \usepackage{steinmetz}

\usepackage[title]{appendix}
\usepackage{algorithm,algpseudocode}
\newcounter{algsubstate}
\renewcommand{\thealgsubstate}{\alph{algsubstate}}
\newenvironment{algsubstates}
  {\setcounter{algsubstate}{0}%
   \renewcommand{\State}{%
     \stepcounter{algsubstate}%
     \Statex {\footnotesize\thealgsubstate:}\space}}

\usepackage{cite}

\usepackage{ifthen}
\usepackage{amsmath}
\usepackage{amssymb}
\usepackage{theorem}
\usepackage{xspace}
\usepackage{calc}
\usepackage{amsfonts}
\usepackage{multicol}
\usepackage{graphicx}	
\usepackage{mwe}
\usepackage{textcomp}
 \usepackage{steinmetz}
 \usepackage{xcolor}

\newcommand{\artworks}{0}
\newcommand{\figures}{1}

\usepackage{ifthen}

\newcommand{\trytoturnpage}{\vspace*{20em}\par\noindent}
\newcommand{\myfig}[1]{\ifthenelse{\artworks=1}{\begin{figure}[f]\trytoturnpage}{\begin{figure}[#1]}}
\newcommand{\mytab}[1]{\ifthenelse{\artworks=1}{\begin{table}[f]\trytoturnpage}{\begin{table}[#1]}}
\newcommand{\myfigstar}[1]{\ifthenelse{\artworks=1}{\begin{figure*}[f]\trytoturnpage}{\begin{figure*}[#1]}}
\newcommand{\mytabstar}[1]{\ifthenelse{\artworks=1}{\begin{table*}[f]\trytoturnpage}{\begin{table*}[#1]}}

\newcommand{\mycaption}[1]{\ifthenelse{\artworks=1}{\vspace*{10em}\caption{#1}}{\caption{#1}}}

\newcommand{\myfigend}{\ifthenelse{\artworks=1}{\trytoturnpage\end{figure}}{\end{figure}}}
\newcommand{\myfigstarend}{\ifthenelse{\artworks=1}{\trytoturnpage\end{figure*}}{\end{figure*}}}

\newcommand{\mytabend}{\ifthenelse{\artworks=1}{\trytoturnpage\end{table}}{\end{table}}}
\newcommand{\mytabstarend}{\ifthenelse{\artworks=1}{\trytoturnpage\end{table*}}{\end{table*}}}

\newcommand{\mycenterwmf}[3]{\ifthenelse{\figures=1}{\centerwmf{#1}{#2}{#3}}{\vskip#2\medskip}}
\newcommand{\myspecial}[1]{\ifthenelse{\figures=1}{\special{#1}}{}}
\newcommand{\mycentereps}[3]{\ifthenelse{\figures=1}{\centereps{#1}{#2}{#3}}{\vskip#2\medskip}}

\newsavebox{\fminibox}
\newlength{\fminilength}
\newenvironment{fminipage}[1][\linewidth]
	{ \setlength{\fminilength}{#1}\addtolength{\fminilength}{-2\fboxsep}%
					       \addtolength{\fminilength}{-2\fboxrule}%
	   \begin{lrbox}{\fminibox}\begin{minipage}{\fminilength}}
	{ \end{minipage}\end{lrbox}\noindent\fbox{\usebox{\fminibox}}}

\newcommand{\gComment}[1]{}

\newcommand{\vet}[1]{{\rm \bf #1}}

{
\theoremheaderfont{\normalfont\scshape}

}

\usepackage{lipsum}

    \newcommand{\gt}{_{GT}}

\newcommand{\begq}{\begin{equation}}
\newcommand{\eeq}{\end{equation}}

\begin{document}
\vspace*{0.2in}

\begin{flushleft}

\newcommand\blfootnote[1]{%
  \begingroup
  \renewcommand\thefootnote{}\footnote{#1}%
  \addtocounter{footnote}{-1}%
  \endgroup
}

{\Large
\textbf\newline{\textbf{Community Detection from  Multiple Observations: 
\\from Product Graph  Model to Brain Applications}}
}

\bigskip

Tiziana Cattai\textsuperscript{1},
Gaetano Scarano\textsuperscript{1},
Marie-Constance Corsi\textsuperscript{2},
Fabrizio De Vico Fallani\textsuperscript{2},
Stefania Colonnese\textsuperscript{1}
\\
\bigskip
\textbf{1} Dept. of Information Engineering, Electronics and Telecommunication, Sapienza University of Rome, Italy 
\\
\textbf{2} Institut du Cerveau et de la Moelle epiniere, ICM, Inserm U 1127, CNRS UMR 7225, Sorbonne Universite, Paris, France
\\
\bigskip

Corresponding author: tiziana.cattai@uniroma1.it

\blfootnote{This work has been submitted to the IEEE for possible publication. Copyright may be transferred without notice, after which this version may no longer be accessible}

\end{flushleft}
\bigskip

\begin{abstract} 
This paper proposes a multilayer graph model for the community detection from multiple observations. This is a very frequent situation, when different estimators are applied to infer graph edges from signals at its nodes, or when different signal measurements are carried out. The multilayer network stacks the graph observations at the different layers, and it links replica nodes at adjacent layers. This configuration matches the Cartesian product between the ground truth graph and a path graph, where the number of nodes corresponds to the number of the observations.  Stemming on the algebraic structure of the Laplacian of the Cartesian multilayer network, we infer a subset of the eigenvectors of the true graph and perform community detection.  Experimental results on synthetic graphs prove the accuracy of the method, which outperforms state-of-the-art  approaches in terms of ability of correctly detecting graph communities. Finally, we show the application of our method to discriminate different brain networks derived from real EEG data collected during motor imagery experiments. We conclude that our approach appears promising in identifying graph communities when multiple observations of the graph are available and it results promising for EEG-based  motor imagery applications.
\end{abstract}

\section{Introduction}
Various cutting-edge fields, such as computational biology, precision medicine, or social studies,   collect measurements over nodes of a network, so  spurring the development of novel tools of graph signal processing (GSP)  \cite{ortega2018graph, sandryhaila2013discrete}. Mining the network community  structure from the observations  boosts analysis  and understanding of the  data \cite{fortunato2010community,lancichinetti2009community}. 

\begin{figure*}[ht!]
	\centering{\includegraphics[scale=0.6]{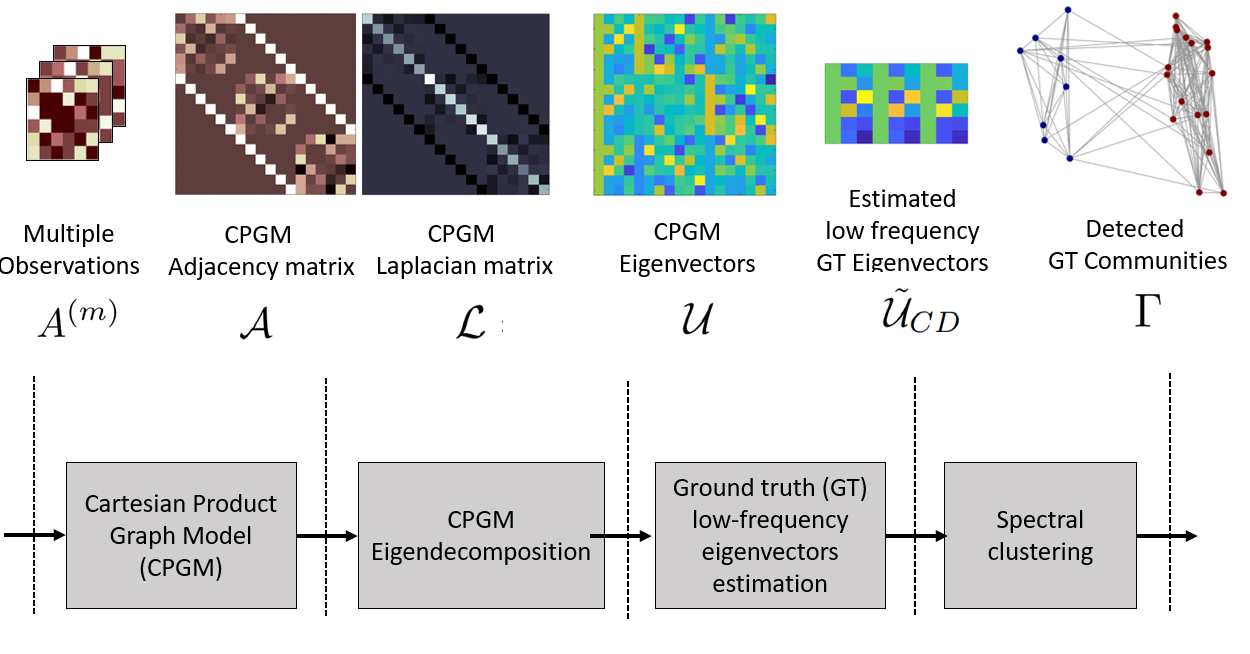}}
\caption{Graphical abstract of the proposed community detection framework.}
    \label{fig:abstract}
\end{figure*}

The human brain stands out as a perfect challenge for GSP-based studies. In this context, researchers represent brain data collected at different EEG electrodes as Signals on Graph (SoG) \cite{ortega2018graph, sandryhaila2013discrete}, and they model the brain structure with graphs. GSP methods can identify network communities and ease the brain network representation \cite{huang2018graph} or the mental state tracking \cite{ting2020detecting}. 
highlighting interactions between different brain regions. EEG-based studies infer interactions  using either several  connectivity  estimators between nodes \cite{bastos2016tutorial} or the same estimator over different frequency bands,   \cite{guillon2017loss}, each leading to different  brain network  representations \cite{presigny2022colloquium}.

Since in this and  many other applications various graph learning methods yield  multiple estimates of  a single unknown ground truth network, a question naturally arises: can we blend the multiple observations to detect the ground truth communities with higher accuracy?  

In this work, i) we introduce a novel model of multiple estimates of the single unknown graph, namely we consider them as different layers of a  multilayer product graph \cite{boccaletti2014structure}, depending on both the unknown ground truth graph  and the number of estimates \cite{cattai2022eeg}; ii) we leverage  the multilayer model to estimate  the ground truth eigenvectors \cite{ortiz2018sampling, sayama2016estimation};  we detect the network communities by  suitably scaling and clustering the ground truth eigenvectors estimated from the multilayer model. 
 
Several state-of-the-art methods  identify communities on multilayer graphs by graph structure analysis \cite{kim2015community,pascal13},   graph matrix factorization  \cite{tang2009clustering,dong2012clustering},    pattern mining functions \cite{zeng2006coherent}, or by specific assumptions on the signal on graph \cite{roddenberry2020exact}. Differently from other state-of-the-art methods \cite{ting2020detecting},  we leverage the  algebraic multilayer graph structure to model the redundancy in  the multiple observations  of the single underlying ground truth graph. This allows us to fully exploit the multiple observations in order to accurately detect the ground truth communities.  Results on synthetic data show that the proposed method outperforms state-of-the-art methods \cite{de2011generalized,blondel2008fast}, \cite{tremblay2014graph,schaub2020blind,roddenberry2020exact}.
Furthermore, we present experimental results on real brain data and show that integrating different graph estimates into the  multilayer model definitely improves the community detection performances.

To sum up, the paper addresses the challenge of community detection given different estimates of a given underlying unknown graph, with a specific emphasis on the analysis of brain data. The main contributions are the following:

\begin{itemize}
    \item we propose to model of multiple and possibly noisy observations of the ground truth graph as a multilayer network;
    \item we adopt a Cartesian Product Graph Model  (CPGM) for the multilayer graph, constructed as the Cartesian product graph between the underlying ground truth graph and a path graph;  we exploit the algebraic CPGM structure to estimate the eigenvectors of the ground truth graph;
    \item we present a novel CPGM based Community Detection (CPGM-CD) algorithm of the ground truth graph, which appropriately scales the estimated first ground truth eigenvectors of the graph, ultimately applying spectral clustering.  
    \item after showing that CPGM-CD  outperforms different state-of-the-art algorithmd, we apply it  to real EEG data recorded during motor imagery-based BCI experiments.  The CPGM, built on several state-of-the-art functional connectivity  estimators, leverages the diversity in deriving the brain network and  captures different  kinds of interactions. We show that CPGM-CD highlights relevant communities, paving the way for deeper understanding of  human brain behaviors. 
 
\end{itemize}
In Fig.\ref{fig:abstract} we present an overview of our method.

The structure of the paper is as follows. Section II describes the observation model. In Section III we introduce the Cartesian Product Graph Model (CPGM) used to merge multiple graph observations. Section IV presents the CPGM- based community detection (CPGM-CD) algorithm. In Section V, we report results on synthetic data, after the description of the generation data model. In Section VI we apply the proposed method to real brain EEG signals, and Section VI concludes the paper.

 In Table \ref{table:not}, we list the main notation used in the paper. 


\begin{table}[b]
 \begin{tabular}{c|c} 
 \hline
 \textbf{Notation} & \textbf{Description}  \\ [0.5ex] 
 \hline\hline
 $\mathcal{A}$,$\mathbf{A}_{GT}$,$\mathbf{A}_{M}$ & adjacency  matrix of multilayer network, ground truth\\ & and path graph\\ 
 \hline

 V & set of nodes at each layer  \\
 \hline
 N & number of nodes at each layer \\
 \hline
 E & set of links at each layer \\ 
 \hline
 K &  number of  links at each layer \\
 \hline
 $\mathcal{L}$,$\mathbf{L}_{GT}$,$\mathbf{L}_{M}$ & Laplacian matrix of multilayer network, ground truth\\ & and path graph\\ 
 \hline
 $\mathcal{U}$,$\mathbf{U}_{GT}$,$\mathbf{U}_{M}$ & Eigenvectors of multilayer network, ground truth\\ & and path graph\\ 
 \hline
M & number of layers \\
\hline
$N_{eig}$ & number of eigenvectors to keep for filtering \\
\hline
$\hat{U}, \Tilde{U}$ & matrix of selected eigenvectors, \\ & normalized matrix of selected eigenvectors\\
 \hline
 P & number of communities \\
 \hline
\end{tabular}
\caption{Table of main notation.}
\label{table:not}
\end{table}

\section{Observation model}
The unknown ground truth graph 
$\mathcal{G}\gt$=($\mathcal{V}\gt$, $\mathcal{E}\gt$) is defined as a set of $N$ nodes (or vertices) and links (or edges). 
The ground truth adjacency matrix $\mathbf{A}\gt \in \mathbb{R}^{N\times N}$ element  $a^{(GT)}_{ij}$ represents  the weight of the link connecting the nodes $i$ and $j$. 
If the ground truth graph $\mathcal{G}_{GT}$ is endowed with a community structure, the vertex set $\mathcal{V}_{GT}$   is partitioned   into $P$ disjoint sets, referred to as node communities. The  $n$-th vertex has an associate to community label $\gamma_n\in\{0,...P-1\}$. The community structure is  represented by a  $P\times N $ affinity matrix $\boldsymbol{\Gamma}$ whose $n$-th column
is a  unit-norm binary vector $\mathbf{1}_p\in \{0,1\}^P$ whose $p$-th component equals to $1$ if the $n$-th node belongs to the $p$-th community and zero otherwise.

 The graph communities  reflect into different network structures  \cite{peel2018multiscale,bassettassortdisassort} based on the conditioning of the affinity matrix $\mathbf{\Gamma}$ on the adjacency matrix $\mathbf{A}$. Herein, we restrict ourselves to the case, by far the most common in the applications, of assortative networks, where nodes belonging to the same community are linked with higher probability than nodes belonging to different communities \cite{newman2002assortative}. Let us remark that widely adopted random network models, such as  the Stochastic Block Models (SBM) and the SBM subset of Planted Partition Models (PPM), are assortative networks \cite{karrer2011stochastic,holland1983stochastic}.

The graph Laplacian $\mathbf{L}\gt\in \mathbb{R}^{N\times N}$ is defined as   
$\mathbf{L}\gt=\mathbf{D}\gt-\mathbf{A}\gt $ 
where $\mathbf{D}\gt$ is the diagonal degree matrix, with diagonal entries $d_i=\sum_{j=1}^{N}a^{(GT)}_{ij}$, $i\!=\!0,...N-1$.
Important features in Graph Signal Processing theory (GSP) are associated to the eigendecomposition of the graph Laplacian  $\mathbf{L}\gt$: 
\begq
\mathbf{L}\gt=\mathbf{U}\gt\mathbf{\Lambda}\gt\mathbf{U}\gt^T
\eeq
 where the diagonal matrix $\mathbf{\Lambda}\gt$ 
  and the orthonormal matrix $\mathbf{U}^{(GT)}$ 
 contains  the Laplacian eigenvalues $\lambda^{(GT)}_n$ and eigenvectors $\mathbf{u}_{GT}^{(n)}$,  $n\!=\!0,...N-1$, respectively. 
 The $n$-th Laplacian eigenvalue  $\lambda^{(GT)}_n$  reflects the properties of the  $n$-th  eigenvector $\mathbf{u}_{GT}^{(n)}$ \cite{cattai2021improving}: eigenvectors with large eigenvalues are  more robust to noise  \cite{milanese2010approximating,michail2018eigenvalues}; eigenvectors with small eigenvalues change smoothly on connected nodes, and   carry information on graph connectivity  \cite{shuman2013emerging, RICAUD2019474}.   

The unknown ground truth graph adjacency matrix $\mathbf{A}_{GT}$ is accessible via $M$   observations,  modeled as follows: 
\begin{equation}
    \mathbf{A}^{(m)} = \mathbf{A}_{GT}+ \mathbf{W}^{(m)}, \;m=0,\cdots M-1
\end{equation}
where  $\mathbf{A}_{GT}$  is a  random matrix  
 drawn from a conditional distribution 
\begq \mathbf{A}_{GT}\approx p_{\mathbf{A}|\mathbf{\Gamma}}(\mathbf{A}|\mathbf{\Gamma}), \;m=0,\cdots M-1\eeq
 and $\mathbf{W}^{(m)}, \;m=0,\cdots M-1$ are i.i.d. random symmetric  matrices independent on $\mathbf{A}_{GT}$ modeling the fluctuations -due to sample noise or estimation errors- of the observed matrices $\mathbf{A}^{(m)}$ with respect to the ground-truth one $\mathbf{A}_{GT}$. 

We leverage this model to develop a community detection algorithm from multiple observations.

\begin{figure}
	\centering{\includegraphics[scale=0.3]{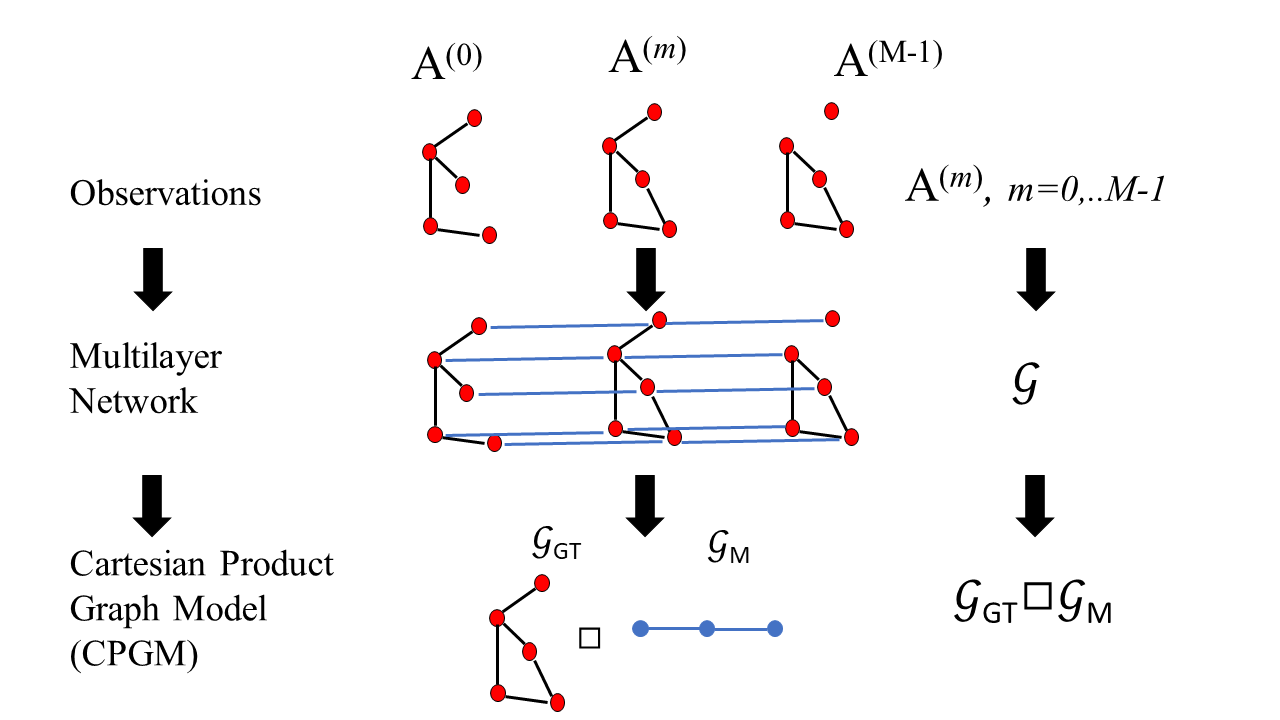}}
\caption{Cartesian product graph model of multiple observations: example of $M=3$ observations of a ground truth graph with $N=5$ nodes (first row), of the  multi-layer network built on  them (second row), and of the  Cartesian product graph model (CPGM) of the observations. The CPGM is obtained from the ground truth graph $\mathcal{G}\gt$ and the $M$ nodes path graph $\mathcal{G}_{M}$. }
    \label{fig:prod_graph}
\end{figure}

\section{From multiple observations to Cartesian Product Graph Model (CPGM)}
State-of-the-art methods may learn the graph from node signals with different criteria, leading to different observed adjacency matrices  \cite{xia2021graph,qiao2018data}.

Herein, we grab the $M$ observations of the graph via a multilayer network model \cite{boccaletti2014structure} defined as follows:
\begin{itemize}
\item the multilayer network contains $M$ layers, of $N$ nodes each;
\item the estimated $m$-th layer adjacency matrix equals the $m$-th estimated adjacency matrix, with $m=0,\cdots M-1$;
\item the corresponding nodes at consecutive layers $(m,m+1),\; m=0\cdots M-2$ are connected.
\end{itemize}
The adjacency matrix $\mathcal{A}$ of the multilayer network is therefore built by $MN\times N$ blocks on the diagonal, representing the links within each layer, as well as by $M-1$ upper-diagonal and $M-1$ lower-diagonal  blocks equal   to the $N\times N$ identity matrix.  
Fig.\ref{fig:prod_graph} (first row) shows the multilayer network model for the case of  $M\!=\!3$ observations of a ground truth graph of $N\!=\!5$ nodes.

Let us remark that for ideal observations, the multilayer network boils down to $M$ replicas of the ground truth graph equipped with links between corresponding nodes.

Our proposed model corresponds to a particular kind of multilayer graph  \cite{chepuri18}, where the network factorizes as the Cartesian product graph of  the ground truth graph $\mathcal{G}\gt$  and a length $M$ path graph $\mathcal{G}_M$:
\begin{equation}
     \mathcal{G}=\mathcal{G}_{GT} \square \mathcal{G}_{M}
\end{equation}
 Therefore, we  model the $M$ observations as layers of a  Cartesian product graph between $\mathcal{G}\gt$ and $\mathcal{G}_{M}$. 
 Fig. \ref{fig:prod_graph} (second row) shows an example of the Cartesian Product Graph Model (CPGM) of multiple observations, where  $M=3$ observations of a ground truth graph with $N=5$ nodes are available (first row). The  multilayer network model (second row) equals the  Cartesian product graph between the  ground truth graph $\mathcal{G}\gt$ and the $M$ nodes path graph $\mathcal{G}_{M}$ (third row).

The Cartesian Product Graph Model (CPGM) endows the observations with a strong algebraic structure. Indeed, the adjacency matrix of the Cartesian product graph $ \mathcal{G}=\mathcal{G}_{GT} \square \mathcal{G}_{M}$, results as the Kronecker sum of the adjacency matrices of ${A}_{GT}$ and ${A}_{M}$:
\begin{equation}
    \mathcal{A}=\mathbf{A}\gt \oplus \mathbf{A}_{M}
\end{equation}
being $\mathbf{A}_{M}$ the adjacency matrix of a path graph of order $M$%
.
The graph Laplacian is obtained as the following Kronecker product:
\begin{equation}
    \mathcal{L}=\mathbf{L}\gt\otimes \mathbf{L}_{M}
\end{equation}

Given  the eigenvectors matrices $\mathbf{U}_{GT},  \mathbf{U}_{M}$ of the ground truth $\mathbf{G}\gt$ and  path graph $\mathcal{G}_{M}$, respectively, the CPGM eigenvectors  are computed as follows \cite{chepuri18}:
 \begq\label{eq:eigprod}
 \mathcal{U}=\mathbf{U}_{M} \otimes \mathbf{U}_{GT}
 \eeq
Besides, the CPGM eigenvalues $\lambda_k,  k\!=\!0,...M\cdot N-1$ are computed by summing up each and every pair of  eigenvalues $\lambda^{(GT)}_n, \;n\!=\!0,\cdots N-1$, and  $\lambda^{(M)}_m,\;m\!=\!0,\cdots M-1$, as follows:  
\begq
 \begin{split}
&\lambda_k\in\left\{\lambda^{(GT)}_n+\lambda^{(M)}_m\right.,\\&
\left.\phantom{\lambda^{(GT)}_n}
m\!=\!0,...M-1, n\!=\!0,...N-1 \right\}
 \end{split}
\eeq

The $NM \times NM$ CPGM eigenvector matrix $\mathcal{U}$ obtained by the Kronecker product of $\mathbf{U}_{M}$ and $\mathbf{U}_{GT}$ is endowed with a strong structure: it stores $M$ differently scaled replicas of the ground truth eigenvectors $\mathbf{u}_{GT}^{(n)}$, with $n\!=\!0,...N-1$, with
 scaling factors equal to the coefficients of the eigenvectors in $U_{M}$. 
 
 Both the eigenvalues $\lambda^{(M)}_m$ and eigenvectors $\mathbf{U}_{M}$ of the length $M$ path graph $\mathcal{G}^{(M)}$ are known in the literature, namely
\begin{equation}
  \lambda^{(M)}_m= 2(1-\cos(\pi m/M))
\end{equation}
and
\begin{equation}
  \mathbf{u}_{M}^{(m)}[l]= \cos(\pi m/M l\!-\!\pi /(2M) l).
\end{equation}

\begin{figure}
	\centering{\includegraphics[scale=0.25]{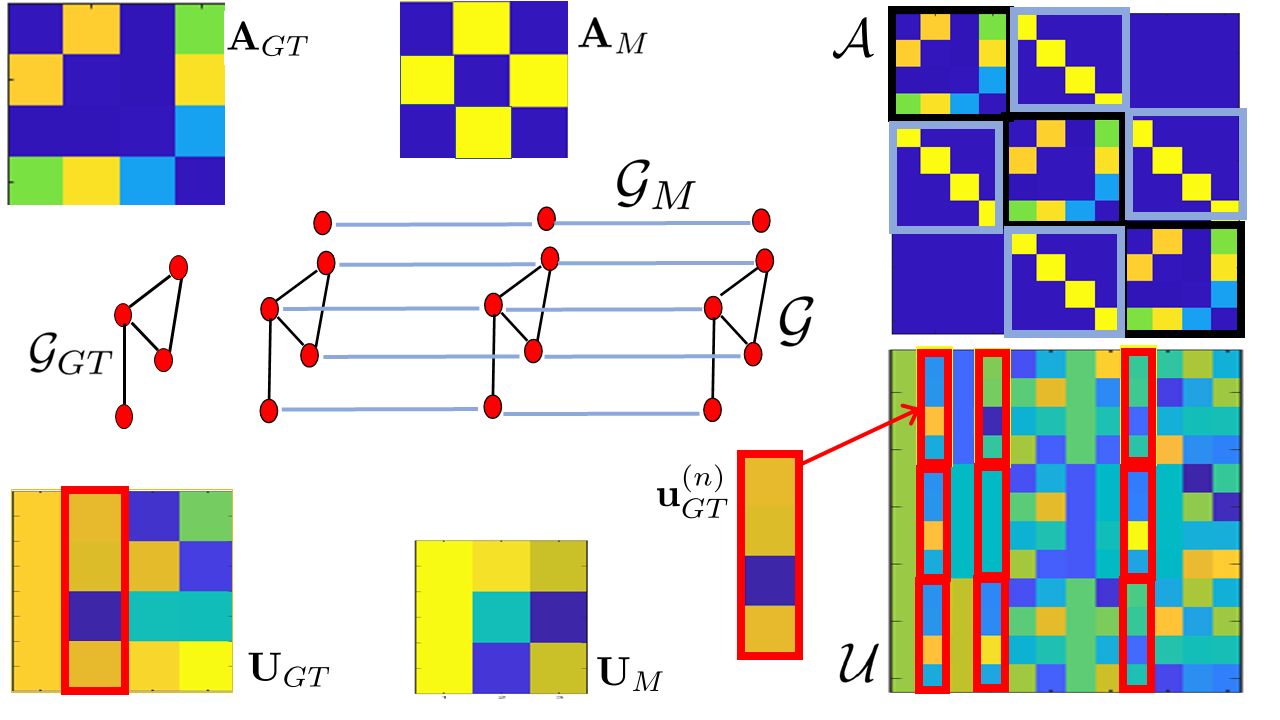}}
\caption{CPGM algebraic structure for an $M=3$ layers' graph $\mathcal{G}=\mathcal{G}_{GT} \square \mathcal{G}_{M}$ of $N = 4$ nodes  each.  
The CPGM adjacency matrix $\mathcal{A}$ is given by $\mathbf{A}\gt \oplus \mathbf{A}_{M}$. The CPGM eigenvector matrix $\mathcal{U}$ is  the  Kronecker product $\mathbf{U}_{GT}\otimes \mathbf{U}_{M}$ of the   path and ground truth graphs' eigenvectors $\mathbf{U}_{GT}$ and $\mathbf{U}_{M}$. Therefore, $\mathcal{U}$ contains $M^2$ replicas of the $n$-th ground truth eigenvector $\mathbf{u}_{GT}^{(n)}, n=0,...N-1$, suitably displaced and scaled. The  scaling factor and position of each replica of $\mathbf{u}_{GT}^{(n)}$ depend on the ground truth eigenvalues and are unknown. }
    \label{fig:sim}
\end{figure}

Fig.\ref{fig:sim} shows a toy example ($M=3$, $N=4$) of the Cartesian product graph $\mathcal{G}$ between the path graph $\mathcal{G}_M$ and the ground truth $\mathcal{G}_{GT}$, of  adjacency  and  eigenvectors matrices  $\mathbf{A}_{M}$, $\mathcal{A}_{GT}$, and  $\mathbf{U}_{GT}$, $ \mathbf{U}_{M}$, respectively. Fig.\ref{fig:sim} also depicts the CPGM adjacency  and the eigenvectors matrices $\mathcal{A}$, $\mathcal{U}$.  The CPGM eigenvector matrix  $\mathcal{U}$ contains $M^2$ replicas of the $n$-th ground truth eigenvector $\mathbf{u}_{GT}^{(n)}, n=0,...N-1$,  displaced and scaled. 

Let us remark that the positions of the replicas of the $n$-th eigenvector 
 $\mathbf{u}_{GT}^{(n)}$ in $\mathcal{U}$  depend  on the ordering of the eigenvalues $\left\{\lambda^{(GT)}_n+\lambda^{(M)}_m,
m\!=\!0,...M-1, n\!=\!0,...N-1 \right\}$, i.e. on the unknown eigenvalues $\lambda^{(GT)}_n$. This 
prevents direct estimation of the eigenvectors $\mathbf{u}_{GT}^{(n)}$ from the multilayer network eigenvectors matrix $\mathcal{U}$.

In the following Section, we propose a community detection (CD) algorithm that leverages the partial knowledge on the  CPGM eigenvector matrix $\mathcal{U}$ to detect ground truth communities.

\section{CPGM based Community Detection (CPGM-CD)}
The CPGM Laplacian $\mathcal{U}$ gathers different replicas of the ground truth eigenvectors $\mathbf{u}_{GT}^{(n)}, n=0,...N\!-\!1$.
We propose a novel CPGM based Community Detection (CPGM-CD) methodology that exploits this algebraic structure by 
\begin{itemize}
    \item[1.]  extracting $N_{CD}<N$  replicas of  selected ground truth  eigenvectors from the CPGM eigenvector matrix $\mathcal{U}$;
    \item [2.]  scaling the replicas to obtain  a  set $\tilde{\mathcal{U}}_{CD}$ of estimated ground truth  eigenvectors;
    \item[3.]  clustering on the estimated eigenvectors' set $\mathcal{U}_{CD}$,  assigning a community label $\gamma_n$ to each node.
\end{itemize} 

\subsection{Extraction Stage}
The first stage extracts the  first $N_{CD}$ eigenvectors 
in $\mathcal{U}$, ordered for ascendant eigenvalues $\lambda_k,  k\!=\!0,...N_{CD}-1$. 
The following $NM\times N_{CD}$ matrix is built:
\begq
\mathcal{U}_{CD}=\left[\mathbf{u}^{(0)}\cdots\mathbf{u}^{(N_{CD}-1)}\right]
\eeq
The first  $N_{CD}$ eigenvectors represent the low frequency components of the Graph Fourier Transform (GFT) basis for the multilayer network, and tend to smoothly vary over connected areas \cite{ortega2018graph}. 

Each   eigenvector $\mathbf{u}^{(k)},k=0,..,N_{CD}$   contains  $M$ replicas of a ground truth eigenvector $\mathbf{u}_{GT}^{(n_k)}$. To determine whether the   ground truth eigenvectors $\mathbf{u}_{GT}^{(n_k)},k=0,..,N_{CD}$ are low frequency or not, we observe that  the CPGM eigenvalues $\lambda_k,  k\!=\!0,...MN-1$ are computed by summing up each and every pair of  eigenvalues $\lambda^{(GT)}_n, \;n\!=\!0,\cdots N-1$, $\lambda^{(M)}_m,\;m\!=\!0,\cdots M-1$. Hence,   the $N_{CD}$ smallest  eigenvalues $\lambda_k,  k\!=\!0,...N_{CD}-1$ correspond to the $N_{CD}$ pairs $(m,n)$ with  the smallest sums  $\lambda^{(GT)}_n+\lambda^{(M)}_m$.  

\begin{figure}
	\centering{\includegraphics[scale=0.25]{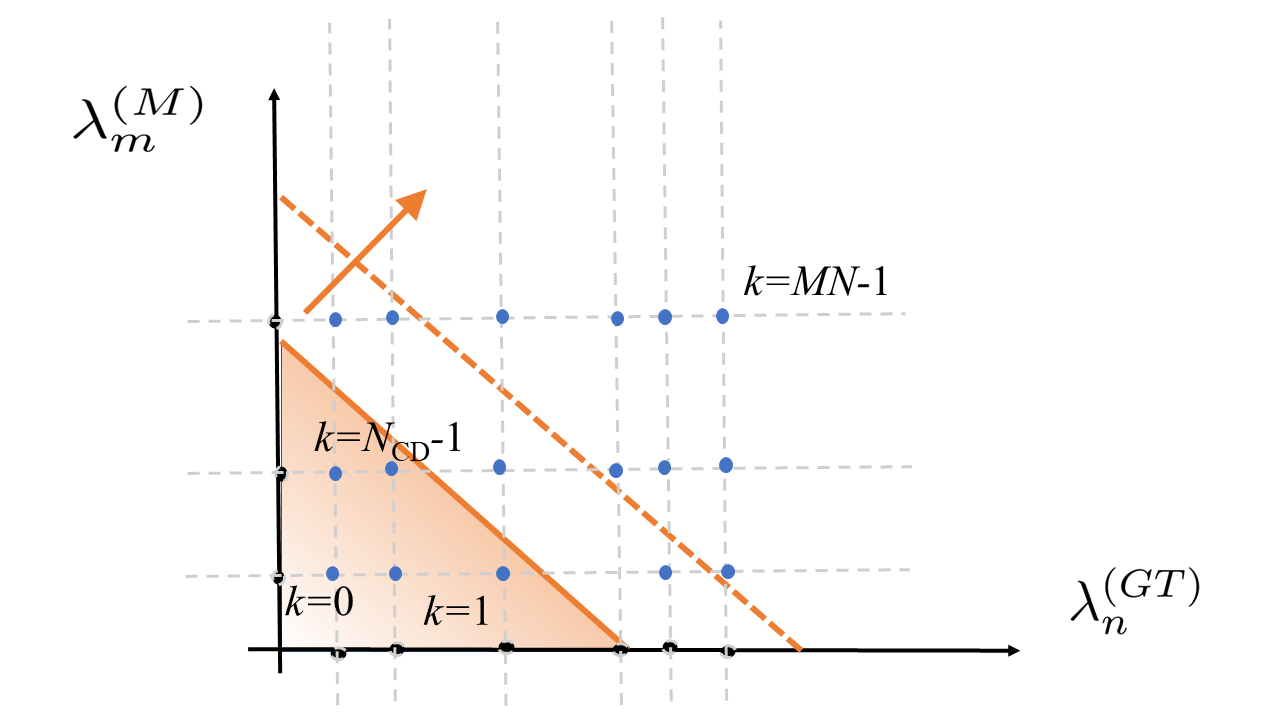}}
\caption{ Example of eigenvalues pairs  $\left(\lambda^{(M)}_m,\lambda^{(GT)}_n\right)$, for $M=3$, $N=6$; the corresponding CPGM eigenvalues $\lambda_k$ are obtained by summing the components $\lambda^{(M)}_m+\lambda^{(GT)}_n $ and are ordered in terms of increasing sum, indicated by the orange slanting lines. The  $\left(\lambda^{(M)}_m,\lambda^{(GT)}_n\right)$ pairs in the orange area contribute to the smallest $N_{CD}$ CPGM eigenvalues, and involve a few of the smallest $N_{CD}$ eigenvalues  $\lambda_{GT}^{(n_k)},k=0,..,N_{CD}$ 
of the unknown ground truth Laplacian $L_{GT}$. }
    \label{fig:eigenval}
\end{figure}

Fig.\ref{fig:eigenval} shows an example of pairs  $\left(\lambda^{(M)}_m,\lambda^{(GT)}_n\right)$, for $M=3$, $N=6$; the corresponding eigenvalues $\lambda_k,  k\!=\!0,...MN-1$ are ordered in terms of increasing sum. The extraction stage selects the eigenvectors corresponding to the smallest $N_{CD}$ CPGM eigenvalues, represented by  the $\left(\lambda^{(M)}_m,\lambda^{(GT)}_n\right)$ pairs in the orange area. The  indexes $n_k,k=0,..,N_{CD}$ of the  ground truth 
eigenvalues $\lambda_{GT}^{(n_k)},k=0,..,N_{CD}$ 
contributing to the smallest $N_{CD}$ eigenvalues $\lambda_k,  k\!=\!0,...N_{CD}-1$ depend  on the unknown ground truth Laplacian $L_{GT}$.  Still, it is clear that the set of $\lambda_{GT}^{(n_k)},k=0,..,N_{CD}$  include some of  the smallest eigenvalues $\lambda^{(GT)}_n$. 
Hence, the  ground truth eigenvectors $\mathbf{u}_{GT}^{(n_k)},k=0,..,N_{CD}$
appearing  in the first $N_{CD}$ eigenvectors of $\mathcal{U}$  
represent low frequency GFT components of the ground truth graph. 

As shown in Fig.\ref{fig:sim}, the matrix $\mathcal{U}_{CD}\in\mathcal{R}^{MN\times N_{CD}}$  of the first $N_{CD}$ eigenvectors $\mathbf{u}^{(k)},k\!=\!0,..,N_{CD}\!-\!1$ collects  $M\times N_{CD}$ scaled replicas
\begq\mathbf{s}_{m}^{(k)}=\alpha^{(n_k)}_{m}\mathbf{u}_{GT}^{(n_k)}\eeq
of the low frequency  ground truth eigenvectors $\mathbf{u}_{GT}^{(n_k)},k\!=\!0,..,N_{CD}$, where the unknown factor 
$\alpha^{(n_k)}_{m}$
accounts for the multiplying coefficient in \eqref{eq:eigprod}.  
 
We denote by ${\mathcal{S}_{CD}}$  the  $N\times M N_{CD}$ matrix defined as 
\\
\begq{\mathcal{S}_{CD}}=
\left[{\mathbf{s}}_0^{(0)}
\cdots 
{\mathbf{s}}_{M-1}^{(N_{CD}-1)}
\right].\eeq

\begin{figure}
	\centering{\includegraphics[scale=0.25]{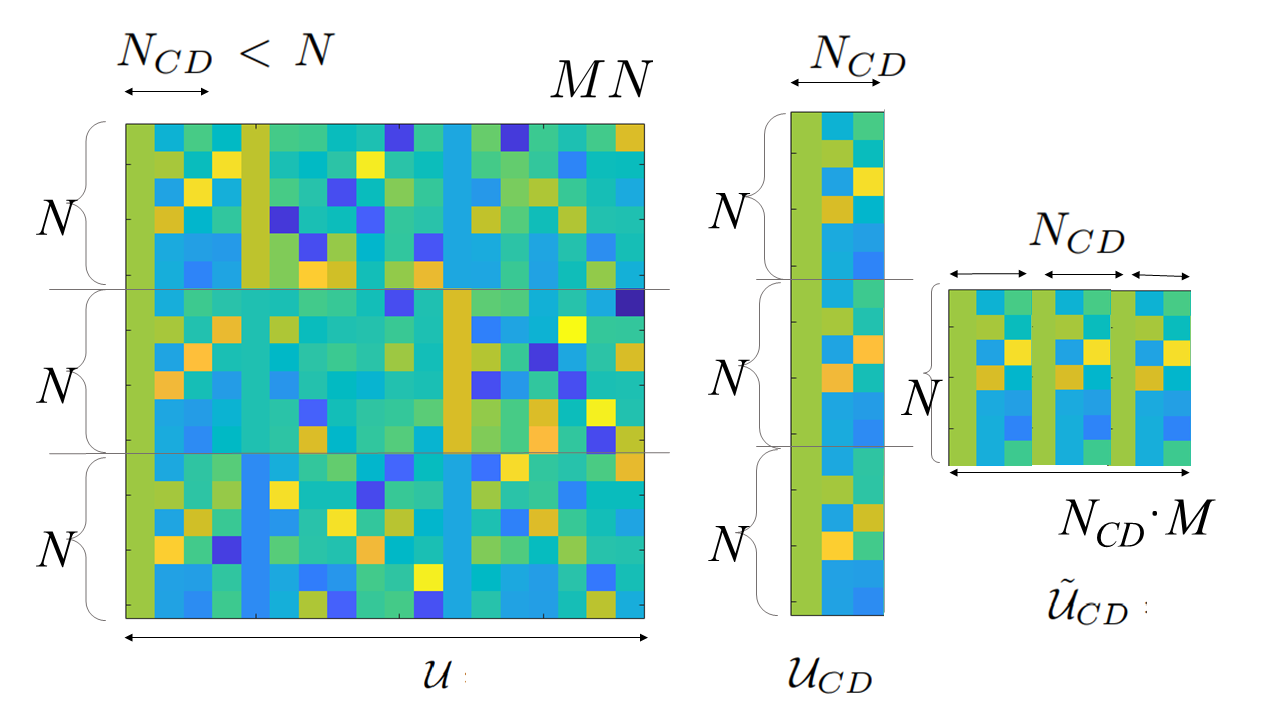}}
\caption{Graphical representation of eigenvectors of the multilayer network with $M=3$ layers and $N=6$ nodes at each layer. In panel A) we report the eigenvector matrix $\mathcal{U}$. In B) we show that we select the first $N_{CD}$ eigenvectors obtaining $\mathcal{U}_{CD}$ with $N\times M=18$, rows and $N_{CD}$ columns. Then, we extract the vectors and arrange them into  the matrix $\tilde{\mathcal{U}_{CD}}$   of dimensions $N\dot N_{CD}M$. 
}
    \label{fig:eig}
\end{figure}

Fig.\ref{fig:eig} depicts, from left to right, the matrices $\mathcal{U} \in \mathcal{R}^{MN\times MN}$, $\mathcal{U}_{CD}\in \mathcal{R}^{MN\times N_{CD}}$ and ${\mathcal{S}_{CD}}\in \mathcal{R}^{N\times M N_{CD}}$  for the toy case of a multilayer network with $M=3$ layers and $N=6$ nodes at each layer. 

We underline that the choice of the parameter $N_{CD}$ determines the number of eigenvectors to be used for the clustering. For this study we leverage results obtained in \cite{cattai2021improving}, and we select the $20 \%$ of eigenvectors associated to largest eigenvalues.

For proper clustering, the column vectors in ${\mathcal{S}_{CD}}$ are scaled as described in the following.

 \subsection{Scaling Stage}
 The columns of the matrix  ${\mathcal{S}}$  are proportional, but not equal,   to the low-frequency ground truth graph eigenvectors. We perform a scaling procedure that i)normalizes the vectors' energy, and ii) flips the vectors so as to align the sign of the zero-th vectors' components. 
 
 This is achieved by computing 
 the normalized matrix for clustering as follows:
 \begq{\tilde{\mathcal{U}}_{CD}}= \mathbf{E}^{-1/2}\cdot\mathbf{B}\cdot
\mathcal{S}_{CD}\eeq
being $\mathbf{E}$ the diagonal energy matrix:
\begq\mathbf{E}= 
\left[
\begin{array}{cccc}
   ||{\mathbf{s}}_0^{(0)}||_2  &0&\cdots &0 \\
   0  &\ddots & \ddots& \vdots\\
   \vdots  &\ddots & \ddots& 0\\
    0 & \cdots&0& ||{\mathbf{s}}_{M-1}^{({N_{CD}\!-\!1})}||_2
\end{array}
\right]
\eeq
and  $\mathbf{B}$ the diagonal binary flipping matrix
\begq\mathbf{B}= 
\left[
\begin{array}{cccc}
   \text{sign}\left(\mathbf{s}_0^{(0)}[0]\right)  &0&\cdots &0 \\
   0  &\ddots & \ddots& \vdots\\
   \vdots  &\ddots & \ddots& 0\\
    0 & \cdots&0& \text{sign}\left(\mathbf{s}_{M-1}^{({N_{CD}\!-\!1})}[0]\right)
\end{array}
\right]
\eeq

For sake of clarity, we underline that the proposed scaling stage acts independently on the clustering. The joint  design of the scaling and clustering steps, although promising, is out of the scope of this paper and we leave it for future work.  

 \subsection{Clustering Stage}
Community detection can be pursued by clustering of the Laplacian eigenvectors, also  known as spectral clustering \cite{white2005spectral, su2019strong}. 
In classical spectral clustering, the $i$-th node of the ground truth graph is associated with the $i$-th row of the eigenvector matrix $U_{GT}$. Clustering is therefore performed over the rows of $U_{GT}$,  and nodes whose rows belong to the  same cluster are assigned the same community label.
Herein, we perform the clustering stage over the $N$ rows of $\tilde{\mathcal{U}}_{CD}$, which contains $N_{CD}$ scaled replicas of the ground truth low frequency eigenvectors, and assign a community label $\gamma_n$ to each node.


To sum up, recent literature on signal-on-graphs has proved that eigenvectors associated to small eigenvalues typically describe the community structure of the graph \cite{ortega2018graph, sandryhaila2013discrete}. With the same strategy, we extend this concept to the multilayer network  by considering the eigenvectors associated to the smallest eigenvalues of the multilayer graph for community detection purposes.
After computing the first $N_{CD}$ eigenvectors of $\mathcal{U}$, we rearrange them in the matrix $\mathcal{S}_{CD}$ of size $N\times M N_{CD}$. Then, we regularize  the energy and direction  of each eigenvector estimate, namely we scale each column of  $\mathcal{S}_{CD}$  to unitary energy and flip the sign such that  its first element is positive. Finally, we identify the communities by spectral clustering of the regularized eigenvectors in $\tilde{\mathcal{U}}_{CD}$  \cite{von2007tutorial},
\cite{shi2000normalized, ng2001spectral}. The spectral clustering provides a  community
 label for each node of the ground truth graph. 
\\
Remarkably, CPGM-CD acts on the multilayer network of $M$ layers representing multiple observations and it identifies the communities of the single underlying ground truth graph. 
The sequence of main steps of CPGM-CD is presented in Algorithm \ref{algo}.

Let us remark that one of the known parameter of our method is the number of graph communities. Indeed, once that we have the estimation of ground-truth eigenvectors $U_{GT}$, we have to perform a spectral clustering with a selected number of clusters to be identified. The number of communities can be estimated leveraging the state-of-the-art on clustering. Indeed many past and recent studies have explored ways to derive the optimal number of clusters for a given data structure, such as the well established consensus clustering \cite{lancichinetti2012consensus}, or with the help of state-of-the-art methods that automatically find the optimal number of clusters  silouette analysis \cite{shahapure2020cluster}. calinski-harabasz index \cite{calinski1974dendrite,wang2019improved}.

\begin{algorithm}[t]
\caption{{\textbf{CPGM-CD Algorithm}}}
\textbf{Input:}  $\mathcal{A}$, $N_{CD}$
 \\
\textbf{Output:}  Community partition ${\Gamma}$  \\
\begin{algorithmic}[1]
\State  Step 1:Eigendecomposition of $\mathcal{L}$
    \begin{algsubstates}
        \State
        Compute the Graph Laplacian of the multilayer network $$\mathcal{L}=\mathcal{D}-\mathcal{A}$$
        \State
        Compute the eigendecomposition
        $$\mathcal{L} = \sum_{q=0}^{N\times M-1}\lambda_q\mathbf{u}_q\mathbf{u}_q^H$$
    \end{algsubstates}
\State Step 2: Community Detection 
    \begin{algsubstates}
        \State
        Select the first $N_{CD}$ eigenvectors of $\mathcal{L}$: ${\mathcal{U}_{CD}}=\left[\mathbf{u}^{(0)}\cdots\mathbf{u}^{(N_{CD}-1)}\right]$
        \State
        Reorganize $\tilde{\mathcal{U}}$ in a strip of N rows and $N_{CD}\times M$ columns
        \State
        Normalize the eigenvectors their  energy and force them to be in the same direction:
        $$\tilde{\mathcal{U}}_{CD}= \mathbf{E}^{-1/2}\cdot\mathbf{B}\cdot \mathcal{S}_{CD}$$
        \State
        $\Gamma$ $\gets$ Spectral clustering $\left(\tilde{\mathcal{U}}_{CD}\right)$
    \end{algsubstates}
\end{algorithmic}
\label{algo}
\end{algorithm}

\section{Experimental results}
Herein, we assess the CPGM-CD method on stochastic graphs, showing that  it outperforms state-of-the-art competitors, especially in  case of  observations affected by large fluctuations.

We generate the $M$ observations in accordance to a Stochastic Block Model, known as  Planted Partition Model  (PPM),  where the set of observable links are related to  the unknown community structure $\Gamma$  as follows.
The $i,j$ element $A_{ij}$ of the ground truth matrix  $\mathbf{A}_{GT}$ is drawn from the following conditional distribution:
\begq 
\begin{split}
&p_{A_{ij}|\gamma_i,\gamma_j}(A_{ij}|\gamma_i,\gamma_j)=\\
&\left\{
\begin{array}{cc}
   \mathbb{P}_s  \delta (A_{ij})+(1- \mathbb{P}_s ) \delta (A_{ij}-1)
    & \gamma_i=\gamma_j \\
   \mathbb{P}_d  \delta (A_{ij})+(1- \mathbb{P}_d ) \delta (A_{ij}-1)   & \gamma_i\neq\gamma_j
\end{array}
\right.  
\end{split}
\label{eq:probmodel}
\eeq
where  the probabilities of observing a link between  nodes of the same or different communities are denoted as $ \mathbb{P}_s $ and $ \mathbb{P}_d $, respectively.

We firstly consider the case where the $M$ observations  differ due to normally distributed estimation error,  i.e.
\begin{equation}
    \mathbf{A}^{(m)} = \mathbf{A}_{GT}+ \mathbf{W}^{(m)}, m=0,...M\!-\!1
\label{eq:normerr}
\end{equation}
where  $\mathbf{W}^{(m)}, m=0,...M\!-\!1$  are i.i.d. Gaussian matrices at   Signal-to-Noise-Ratio SNR%
\footnote{Specifically,
$w_{ij}^{(m)}\!=\!w_{ji}^{(m)}\approx
\mathcal{N}(0, \sigma_w^2/2)$, where $$\sigma_w^2=\dfrac{1}{N^2}|| \mathbf{A}_{GT}||_F^2\cdot\text{SNR}^{-1}$$ and $||\cdot ||_F$ denotes the matrix Frobenius norm.}.

\begin{figure}
 \includegraphics[scale=0.4]{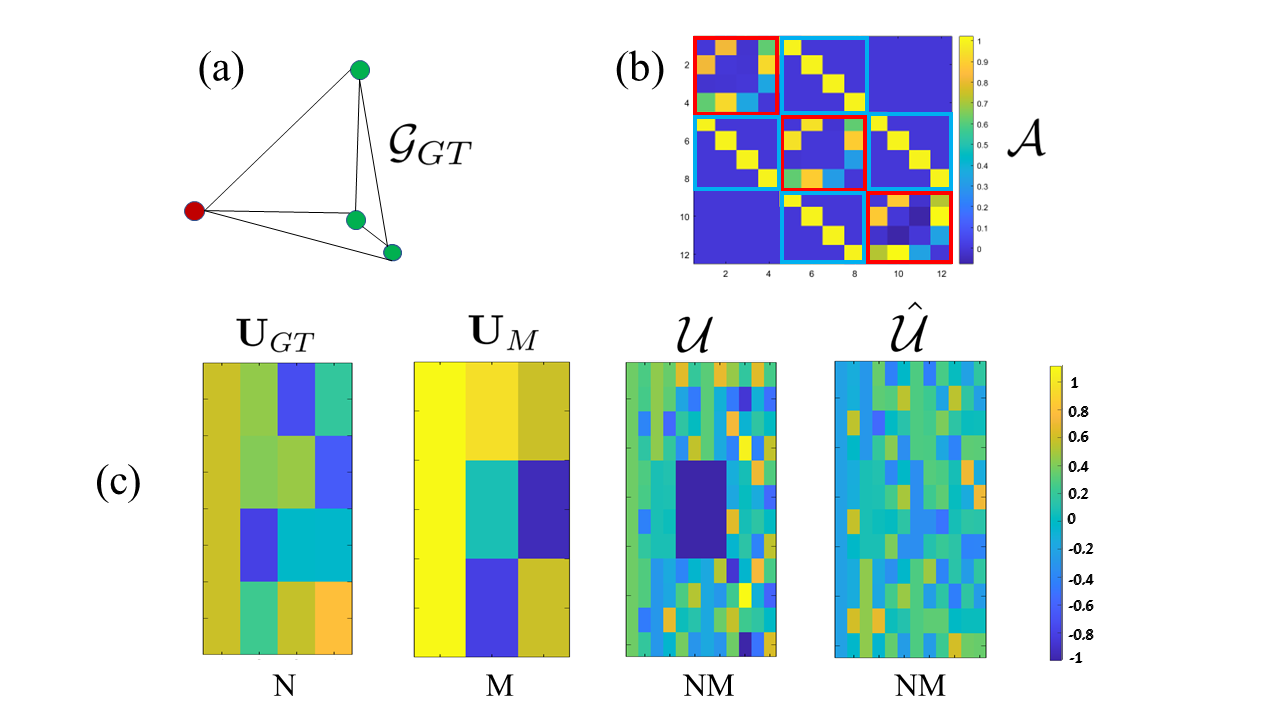}
\caption{(a) Graph with $N = 4$ nodes belonging to   $P=2$ communities generated according to \eqref{eq:probmodel}, \eqref{eq:normerr} with $p_s =1-p_d = 0.9$  and $\text{SNR}=40$ db; (b) multilayer adjacency matrix $\mathcal{A}$ built on  $M=3$ observations $\mathbf{A}^{(m)}$; (c) ground truth and  path graph eigenvector matrices $U_{GT}$, $U_{M}$, and ground truth and estimated multilayer eigenvector matrices $\mathcal{U}= \mathbf{U}_{GT}\otimes \mathbf{U}_{P}$, $\hat{\mathcal{U}}$. The CPGM-CD  algorithm exactly recovers the  community labels. }
    \label{fig:synthres}
\end{figure}

 Fig. \ref{fig:synthres} refers to the toy case of a ground truth graph  of $N = 4$ nodes belonging to  $P=2$ communities; the graph, generated according to the PPM model using $p_{s} \!=\! 0.9,\;p_{d} \!=\! 0.1$ is sketched in Fig. \ref{fig:synthres}(a).  The $M=3$ observations $\mathbf{A}^{(m)}, m=0,1,2$ are affected by an additive Gaussian noise at   $\text{SNR} \!=\! 40$ dB. A multilayer network  formed  by $M\!=\!3$ layers with $N\!=\!4$ nodes  each is built;   Fig. \ref{fig:synthres}(b) plots the related  multilayer adjacency matrix $\mathcal{A}$, where the diagonal blocks  represent the observations $\mathbf{A}^{(m)}, m=0,1,2$, and the upper and lower diagonal blocks are identity matrices representing the links between corresponding nodes on adjacent layers. The algebraic structure of the  multilayer adjacency matrix $\mathcal{A}$ reflects into the structure of the multilayer eigenvector matrix $\mathcal{U}$. 
 
Fig. \ref{fig:synthres} (c) represents 
the $4\times4$ ground truth eigenvector matrices $\mathbf{U}_{GT}$, 
the $3\times3$ path graph eigenvector matrices $\mathbf{U}_{M}$,   the $12\times12$ multilayer ground truth  eigenvector matrices $\mathbf{U}$ -obtained as the Kronecker product of $\mathbf{U}_{GT}$ and $\mathbf{U}_{M}$-,  and the estimated multilayer eigenvector matrix $\hat{\mathbf{U}}$ -obtained by eigenanalysis of the  multilayer Laplacian $\mathcal{L}$ obtained from $\mathcal{A}$-.
The coefficients of the first estimated eigenvectors in $\hat{\mathbf{U}}$ in Fig. \ref{fig:synthres}(c), apart from the different ordering  and from the energy normalization which reflect into different colors in the plot,  match the first eigenvectors based on the CP model in $\hat{\mathbf{U}}$. These vectors are selected and reshaped to apply the scaling and clustering stages of the CPGM-CD algorithm, that correctly recovers the nodes community labels $\gamma_i$, $i=0,\dots 3$. 

\begin{figure}
	\centering{\includegraphics[scale=0.6]{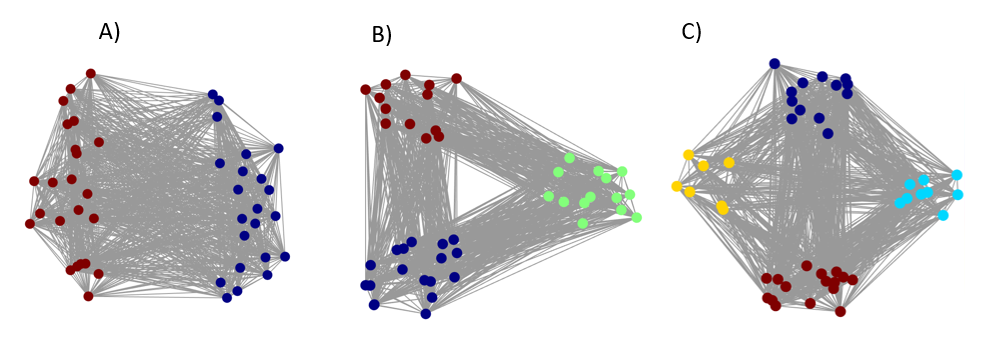}}
\caption{Application of the CPGM-CD method  to the case of $M\!=3\!$ observations of a  $N\!=\!45$ nodes graph with $P\!=\!2,3,4$ communities ($p_s\!=\!1-p_d\!=\! 0.8$ and $\text{SNR} \!=\! 40$dB). We represent the community labels  recovered by the CPGM-CD method in terms of colors, where each color represents a specific community. The estimated communities perfectly match the true one.}
    \label{fig:3com}
\end{figure}

 Fig. \ref{fig:3com} plots the result of the community detection algorithm on three PPM graphs of $N=45$ nodes, belonging to $P=2,3,4$ communities, and generated according  $p_{s}\! =\! 0.8$, $p_{d}\!=\! 0.2$. The $M=3$ observations are obtained at $\text{SNR} \!=\! 40$.  The CPGM-CD algorithm is applied by leveraging $N_{CD}\!=\!3$ eigenvectors, of $NM\!=\!45\cdot 3$ coefficients each, suitably reshaped  into the $N\times M\cdot N_{CD}=45\times 9$ matrix $\mathcal{S}_{CD}$. After proper scaling and clustering, the CPGM-CD method correctly recovers all the community labels, identified by different colors  in Fig.\ref{fig:3com}.

\begin{algorithm}[ht!]
\caption{\textbf{Majority voting algorithm}}
\textbf{Input:}  $\Gamma^{(m)}$
 \\
\textbf{Output:}  Community partition ${\Gamma}$, P  \\
\begin{algorithmic}[1]
\State Majority Voting
     \For{{i=1:N}}
        \State {
        \For{{p=1:P}}
        \State {
        votes(p)=sum((logical($\gamma_i$==p)));
        }
        \EndFor
        \State
        $max_{majvote}$ $\gets$ max(votes)
        \State
        $index_{majvote}$ $\gets$ find(votes == $max_{majvote}$)
        \State
        $index_{random}$ $\gets$ randomization of $index_{majvote}$
        \State
        $\gamma_i$ $\gets$ $index_{majvote}(index_{random})$
       }
      \EndFor

\end{algorithmic}
\end{algorithm}

We now compare the  CPGM-CD  performances with  those achieved by  i)separately applying state-of-the-art community detection algorithms to $\mathbf{A}^{(m)}, m\!=\!0,...M-1$ and ii) assigning   the community  to each node  by  majority voting after suitable harmonization of the labels resulting at different layers. 

We considered  different single-layer community detection methods, namely the Spectral Graph Wavelet Transform based multiscale method in \cite{tremblay2014graph},   the Louvain Algorithm \cite{blondel2008fast}, as well as Spectral Clustering. 
 The three methods obtained by cascading each  single-layer method with the   label harmonization and majority voting are hereafter referred to as Multi-Layer SGWT (ML-SGWT) Multi-layer Louvain Algorithm (ML-LA) and Multi-Layer Spectral Clustering (ML-SC), respectively.

In addition, we consider a community detection method specifically introduced for multilayer networks in \cite{roddenberry2020exact}. According to the method introduced by the authors we derive the signal on graph:
$\vet{y} = \mathcal{H}(S)\vet{w}$, where $\vet{w}$ is a white excitation drawn from the normal distribution  $\mathcal{N}(0,I_N)$. With this state, the estimated covariance matrix $\hat{K}_{yy}$ results
$\hat{K}_{yy}=\dfrac{1}{M}\sum\limits_0^{M}\vet{y}\vet{y}^T$. We consider a network process represented by the graph filter 
$H(L(m)) = (I-\beta L(m))^5$, with $\beta = 1/(4 + 4\gamma) log N$, $\gamma=0.3$.


We evaluate the community detection performance by computing the Normalized Mutual Information (NMI) between the detected and  ground truth community labels \cite{vinh2009information}; therefore, the NMI represents the similarity of the estimated communities to the real ones.

\begin{table}[t]
\centering
 \begin{tabular}{|l|c|c|c|c|c|}
 \hline
   \textbf{NMI  vs SNR } & $50$dB & $40$dB & $30$dB & $20$dB & $10$dB  \\ 
 [0.5ex] 
 \hline
ML-LA \cite{blondel2008fast} & $\mathbf{1}$  & 0.49 & 0.97 & 0.18 & 0.04  \\ [0.5ex]  
 \hline  
 ML-SC \cite{ng2001spectral}& 0.87 & 0.24 & 0.47 & 0.33 & $\mathbf{0.07}$ \\ [0.5ex] 
  \hline  
 ML-SGWT ($s\!=\!1$)  \cite{tremblay2014graph}
 & 0.03 & 0.02 & 0.02& 0.03 & 0.03\\ [0.5ex] 
 ML-SGWT ($s\!=\!2$) \cite{tremblay2014graph} & 0.02 & 0.03 & 0.03& 0.03 & 0.03 \\ [0.5ex] 
 ML-SGWT ($s\!=\!3$) \cite{tremblay2014graph} & 0.02 & 0.05 & 0.09& 0.03 & 0.04 \\ [0.5ex] 
 ML-SGWT ($s\!=\!4$) \cite{tremblay2014graph} &$\mathbf{1}$ & 0.02 & 0.13 & 0.06 & 0.04 \\ [0.5ex] 
ML-SGWT ($s\!=\!5$) \cite{tremblay2014graph} & $\mathbf{1}$&  0.86 & 0.62 & 0.17 & 0.05 \\ [0.5ex] 
ML-SGWT ($s\!=\!6$) \cite{tremblay2014graph}  & $\mathbf{1}$& $\mathbf{1}$ & 0.56 & 0.19 & 0.05 \\ [0.5ex] 
\hline  
Exact Blind \cite{roddenberry2020exact}  & 0.03 & 0.03 & 0.03 & 0.03 & 0.03 \\ [0.5ex] 
\hline 
Exact Blind (M $\rightarrow {\infty}$) \cite{roddenberry2020exact}  & $\mathbf{1}$& 0.52 & 0.36 & 0.27 & 0.05\\ [0.5ex] 
\hline 
 \textbf{CPGM-CD}  & $\mathbf{1}$ & $\mathbf{1}$ & $\mathbf{1}$ & $\mathbf{0.39}$ & $\mathbf{0.07}$ \\ [0.5ex] 
  \hline  
\end{tabular}

\caption{Normalized Mutual Information (NMI) between the ground truth  and estimated community labels for the CPGM-CD method and for state-of-the-art community detection methods, for a PPM graph as in in \eqref{eq:normerr} ($N\! =\!30$, $P\! =\!2$, $M\! =\!3$,  $ \mathbb{P}_s \! =\!1- \mathbb{P}_d \! =\!0.8$); the  NMI is averaged over $500$ runs.  }
\label{table:NMI_samea}
\end{table}



In the first case, we refer to the observation model in \eqref{eq:normerr}, where the $M$ observed adjacency matrices differ due to a Gaussian estimation error.  We consider different Signal-to-Noise-Ratio (SNR) levels, namely $50,40,30,20,10$ dB. For this simulation, the NMI is averaged over $500$ runs to increase the statistical validity.   Table \ref{table:NMI_samea} reports the NMI value associated to the CPGM-CD  method and to the  competitors. For the multiscale ML-SGWT method, we report the results for different values of the scale parameter $s$.  The CPGM-CD systematically achieves the highest NMI values with respect to all the alternatives, not only at  high SNR  but also at low SNR. Let us remark that for Spectral Graph Wavelet Transform method \cite{tremblay2014graph}, we consider the un-normalized graph Laplacian because it is more robust to noise. The normalized graph Laplacian enables better NMI values for $50,40,30$ dB while it breaks for less SNR values. Thereby, in the case of noisy observations, the graph algebraic  structure underlying the CPGM-CD effectively counteracts the error in the estimation of the adjacency matrices.


We now evaluate the performance of the CPGM-CD under the following  observation model:
\begin{equation}
    \mathbf{A}^{(m)} = \mathbf{A}^{(m)}_{GT}+ \mathbf{W}^{(m)}, m=0,...M\!-\!1
\label{eq:samperr}
\end{equation}
where  $M$ ground truth adjacency matrices $\mathbf{A}^{(m)}_GT$ are independently and randomly drawn under the same community partition $\Gamma$ and observed in presence of i.i.d additive Gaussian noise $\mathbf{W}^{(m)}, m=0,...M\!-\!1$. Thereby,  the observed adjacency matrices $\mathbf{A}^{(m)}$ share the ground-truth community structure  $\Gamma$ but they differ  even at high SNR, due to sample noise. 
This scenario is realistic  because it captures the statistical variability of real networked data, which are often taken in different experimental trials and therefore may present  different network structures even though they share the same community structure. This scenario is also  challenging  because the multilayer graph built using the observed adjacency matrices deviates   from the Cartesian Product model underlying CPGM-CD even in the noise free case.

We report the NMI achieved by the different methods in Table \ref{table:NMI_samegamma},  for the case of $N =30$, $P=2$, $M=3$ with $500$ repetitions in order to improve the statistical validity of the simulations, at  $\textbf{SNR}=Noise-free, 40,15,5,-5$.
We recognize that also in this realistic and challenging scenario CPGM-CD achieves the maximum NMI on a wide range of SNR levels.

\begin{table}[t]
\centering
 \begin{tabular}{|l|c|c|c|c|c|}
 \hline
   \textbf{NMI  vs SNR } & Noise-free & $40$dB & $15$dB & $5$dB & $-5$dB  \\ 
 [0.5ex] 
 \hline
ML-LA \cite{blondel2008fast} & 0.99  & 0.99 & 0.98 & 0.86 & 0.15\\ [0.5ex]  
 \hline  
 ML-SC \cite{ng2001spectral}& 0.97  & 0.97& 0.89 & 0.72 & 0.19  \\ [0.5ex] 
  \hline  
 ML-SGWT ($s\!=\!1$)  \cite{tremblay2014graph}
 &0.03 &  0.02 & 0.02 & 0.03& 0.03\\ [0.5ex] 
 ML-SGWT ($s\!=\!2$) \cite{tremblay2014graph} & 0.02 & 0.03  & 0.03 & 0.03 & 0.03   \\ [0.5ex] 
 ML-SGWT ($s\!=\!3$) \cite{tremblay2014graph} & 0.05 & 0.05 & 0.06 & 0.07 & 0.05 \\ [0.5ex] 
 ML-SGWT ($s\!=\!4$) \cite{tremblay2014graph} & 0.43  & 0.47 & 0.25 & 0.16 & 0.07 \\ [0.5ex] 
ML-SGWT ($s\!=\!5$) \cite{tremblay2014graph} & 0.95 & 0.94 & 0.83 & 0.45 & 0.13  \\ [0.5ex] 
ML-SGWT ($s\!=\!6$) \cite{tremblay2014graph} & 0.98 & 0.97 & 0.88 & 0.59  & 0.13\\ [0.5ex] 
\hline  
Exact Blind \cite{roddenberry2020exact}  & 0.03  & 0.03 &  0.03 & 0.03 & 0.03 \\ [0.5ex] 
\hline 
Exact Blind (M $\rightarrow {\infty}$)  & 0.94 &  0.94& 0.93 & 0.79 & 0.15\\\cite{roddenberry2020exact}  &  &  &  &  &  \\ [0.5ex] 
\hline 
 \textbf{CPGM-CD}  & $\mathbf{1}$ & $\mathbf{1}$  & $\mathbf{1}$  & $\mathbf{0.99}$  & $\mathbf{0.5}$  \\ [0.5ex] 
  \hline  
\end{tabular}

\caption{Normalized Mutual Information (NMI) between the ground truth  and estimated community labels for the CPGM-CD method and for state-of-the-art community detection methods,  for a PPM graph as in \eqref{eq:samperr} ($N\! =\!30$, $P\! =\!2$, $M\! =\!3$,  $p_s\! =\!1-p_d\! =\!0.8$); the  NMI is averaged over $500$ runs. 
}
\label{table:NMI_samegamma}
\end{table}

The results in Tables \ref{table:NMI_samea},\ref{table:NMI_samegamma} clearly show that the CPGM-CD accurately detects the graph communities and  outperforms  state-of-the-art alternatives, leveraging  a Cartesian Product  model of the observations     obtained by different  estimators and/or in different trials. 

A few remarks are in order.  In a nutshell, the performance  of CPGM actually depend on  how much the fluctuations of the observations $\mathbf{A}^{(m)}$ 
affect the low-frequency subspace of the multi-layer eigenvector matrix $\mathcal{U}$.  This depend on  network specific factors, namely i) the random  variability of the $N\times N$ graph edges, which result in a population noise depending on the network size $N$, and ii) the network ground-truth connectivity, that affects the robustness of the multi-layer low-frequency subspace. Further work is needed to  relate the accuracy  of CPGM-CD  to the  ground-truth network structure (e.g. $N$, $P$, $\boldsymbol{\Gamma}$ as well to the observation model as in \eqref{eq:normerr} and \eqref{eq:samperr}. We summarize here  the main trends stemming from the experiments,  leaving the analytical performance evaluation for future study.
Firstly, for given PPM parameters $p_s,p_d$, large network structures (large $N$) give rise to a larger number of observed links, leading to better accuracy. Secondly,  the  number of estimated eigenvectors $\tilde{\mathcal{U}}_{CD}$ at the input of the clustering stage depends on the product $M\cdot N_{CD}$; therefore,  a larger number of observations $M$ can be traded with a smaller number of multilayer eigenvectors $N_{CD}$. This condition is favorable since the subspace of the low-frequency eigenvectors in $\mathcal{U}$ is robust to noise \cite{cattai21sipn}. Finally,  when the number of communities $P$ is small compared to $N$, the community partition is easier highlighted by the given number of  observed links.

\section{Application on real BCI experiments}
Here, we present experimental results of our proposed community detection algorithm on real  data, recorded during motor-imagery BCI experiments. 
In particular, we use open EEG data related to the BCI competition IV dataset $1$ \cite{tangermann2012review}, where a set of  healthy subjects participated in the study. 
The subjects performed a motor imagery tasks without feedback. For our study we consider motor imagery of the right hand and resting state with total of $100$ trials per condition. 

Our study represents a proof-of-concept of the applicability of the proposed community detection to brain data. For sake of simplicity, we consider just one subject. 

We now  apply the CPGM-CD method  in this framework, in order to   identify  communities associated to the motor-imagery tasks. 

\subsection{Building the multilayer network with real EEG data}
The first step consists in building the multilayer network from the original EEG data. 
For each mental state, $N=59$ EEG time series  at $100$ Hz sampling rate are obtained at  the electrodes  locations \cite{tangermann2012review}.
We   compute different functional connectivity (FC) estimators  on the EEG signals at  nodes' pairs \cite{bastos2016tutorial}. Since  each FC estimator measures the statistical interaction between brain areas in a different way,    the resulting FC graph representations differ.

In this work, we apply three classical FC estimators, i.e. the spectral coherence \cite{carter1987coherence}, the imaginary coherence \cite{nolte2004identifying} and the phase difference \cite{cattai2021phase}, to characterize and detect brain  states during motor imagery tasks  \cite{cattai2021phase}. 
\\
Let us consider 
the EEG time series collected at the graph vertices $i$ and $j$ 
 and  let us denote by $\hat{P}_i(\omega_k)$, $\hat{P}_j(\omega_k)$ and $\hat{P}_{ij}(\omega_k)$ their auto-spectra and cross-spectrum at the frequencies   $\omega_k, \;{k=0},...{K-1}$, respectively. For each nodes pair $i,j$, the spectral coherence $C_{ij}$ \cite{carter1987coherence}, the imaginary coherence $IC_{ij} $\cite{nolte2004identifying} and the phase difference $ \Phi_{ij}$ \cite{cattai2021phase}  are computed as the following spectral averages: 
\begin{equation}
    C_{ij}=\dfrac{1}{K}
      \sum\limits_{k\!=\!0}^{K\!-\!1}
    \big| \hat{P}_{ij}(\omega_k)\big|
    \;\big(\hat{P}_i(\omega_k)\!\cdot\! \hat{P}_j(\omega_k)\big)^{-1},
    \label{eq:coh}
\end{equation}

\begin{equation}
    IC_{ij}=\dfrac{1}{K}
      \sum\limits_{k\!=\!0}^{K\!-\!1}\left| \text{Im}\big\{\hat{P}_{ij}(\omega_k)\big\}\right|
 \;\big(\hat{P}_i(\omega_k)\!\cdot\! \hat{P}_j(\omega_k)\big)^{-1},    \label{eq:imagcoh}
\end{equation}

\begin{equation}
    \Phi_{ij}=\dfrac{1}{K}
      \sum\limits_{k\!=\!0}^{K\!-\!1}\vert{\phase{\hat{P}_i(\omega_k)}-\phase{\hat{P}_j(\omega_k)}\vert}.
\end{equation}

We consider $K=14$ frequencies $\omega_k, $ corresponding to equispaced points in the  $\alpha$ frequency band ($8\!-\!13$ Hz) involved by motor imagery tasks \cite{pfurtscheller2001motor,pfurtscheller2006mu}. 
 
 \begin{figure}[h!]
	\centering{\includegraphics[scale=0.57]{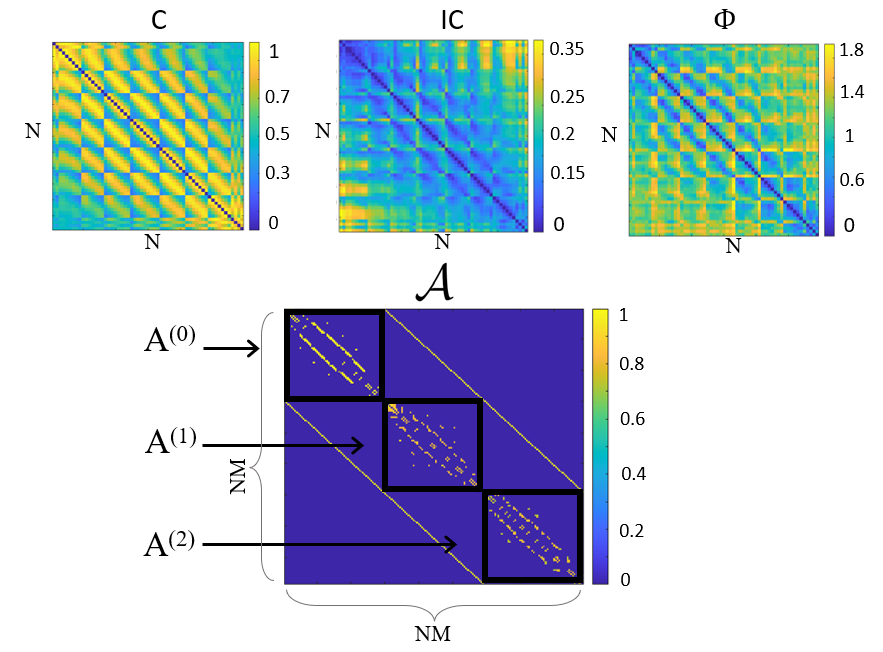}} 
\caption{Top line: Three FC connectivity estimators used for the model, i.e. coherence, imaginary coherence, phase difference. Bottom line: the adjacency matrix of the multilayer network, after  rescaling and thesholding of each FC estimator to keep $5\%$ of its links. In this case, we consider one trial of resting state. In the principal diagonal, we have $A^{(0)}$,$A^{(1)}$,$A^{(2)}$ derived from the FC computation with the three estimators.}
    \label{fig:ml_real}
\end{figure}

 In the top row of Fig.\ref{fig:ml_real} we represent the three matrices for an example of the first trial of resting state. It is possible to remark that the meaning and the scales change according to the FC estimator used for the analysis. In order to apply the Cartesian Product Graph Model, we threshold the matrices to keep a fixed percentage of the strongest connections, and  rescale the values as follows. Let us remark that other methods have been recently proposed to threshold the adjacency matrix \cite{de2017topological}, but we decided for the easiest and most intuitive solution.

The three observations $\mathbf{A}^{(m)},m\!=\!0,1,2$ are obtained from the different FC estimators after recasting the values in the range $[0,1]$, with $1$ corresponding to maximally  correlated signals. This step is necessary in order to have coherent connectivity estimates, which corresponds to have multiple graph observations according to our Cartesian Product Graph Model. Specifically, for any $i,j$, $i \neq j$,  we set  $${A}^{(0)}_{ij}\!=\!{C}_{ij}, \;{A}^{(1)}\!=\!1-{IC}_{ij}/M_{IC},\;{A}^{(2)}_{ij}\!=\!1-{\Phi}_{ij}/M_{\Phi},$$ being $M_{IC}\!=\!\underset{ij}{\max}\;{IC}_{ij}$ and $M_{\Phi}\!=\!\underset{ij}{\max}\;{\Phi}_{ij}$; besides, ${A}^{(m)}_{ii}\!=\!0, m\!=\!0,...M\!-\!1,i\!=\!0,...N\!-\!1,$.  

 Fig.\ref{fig:ml_real} shows in the bottom line, the   multilayer adjacency matrix $\mathcal{A}$ after thresholding and rescaling  the original FC estimators, and we can see  the three observed matrices $\mathbf{A}^{(m)},m\!=\!0,1,2$  on the principal block diagonal and the upper and lower block diagonals are the identity matrices, corresponding to the interlayer connections.

\subsection{Community detection of the multilayer network with real EEG data}
In this subsection, we apply our proposed community detection algorithm to real EEG data previously described.  In order to investigate the community organization during different human mental states, we consider EEG data recorded one subject performing two different tasks, that are: motor-imagery of the right hand  and resting state. This study represents a proof of concept applied on single subject to  show the potential of our method to identify the community structure on brain data. Interestingly, the analysis can be extended to a large number of subject to evaluate the impact of the estimated  community structure in a classification system such as a Brain Computer interface; this is left for further study.

\begin{figure}[h!]
    \centering
    \includegraphics[scale=0.9]{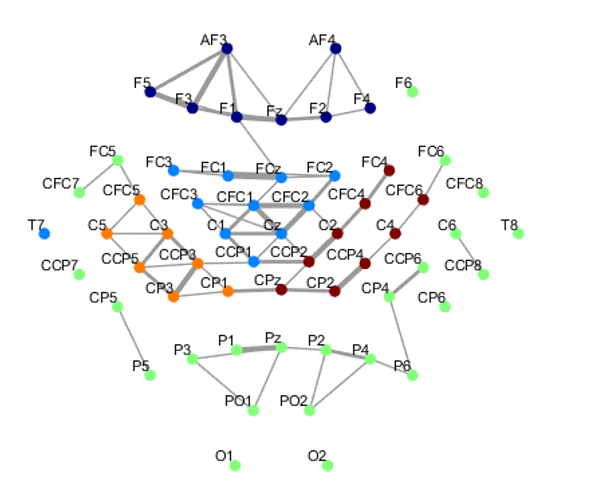}
    \caption{Result of community detection algorithm in terms of communities represented on graph for resting state in $\alpha$ band for all the $100$ trials. The colour of the node represents the community to whom it belongs. The percentage of links kept of this analysis is $2\%$ }
    \label{fig:CD_graph}
\end{figure}

In order to understand the functioning of the method, we present in Fig. \ref{fig:CD_graph} the result of the community detection algorithm on resting state. In particular, we consider FC estimators computed in $\alpha$ frequency band and we replicate the analysis for the $100$ available trials of resting state. In order to have a compact but representative visualization, we derive the most representative trial. 
To this aim, we compute the NMI for each pair of trials, since it measures the similarity between communities obtained at different observations. Then, for each trials, we sum all the derived NMIs and we take the one associated to the largest sum. This operation identifies the most representative trial because it is the most similar to all the others.
Results are reported in terms of colors, which means that each node of the network is coloured according to the  community identified by the algorithm. In this case, $P=5$.
In Fig. \ref{fig:CD_graph} the link  width represents the stability between the three networks: indeed, the link is widest when it is present (after thresholding) on the three networks and it is absent if there is no link in any graph.
For what concerns the connections, we recognize that the  stable links between nodes, represented by thick gray line,  are captured by the multilayer mode, and the CPGM-CD detects  communities, represented by nodes of the same colors, that reflect such stable links. More specifically, we recognize links between frontal nodes (in the blue community), links between parietal nodes (green community) and links in left, right and central areas, respectively in  orange, red and light blue. 

\begin{figure*}
	\centering{\includegraphics[scale=0.75]{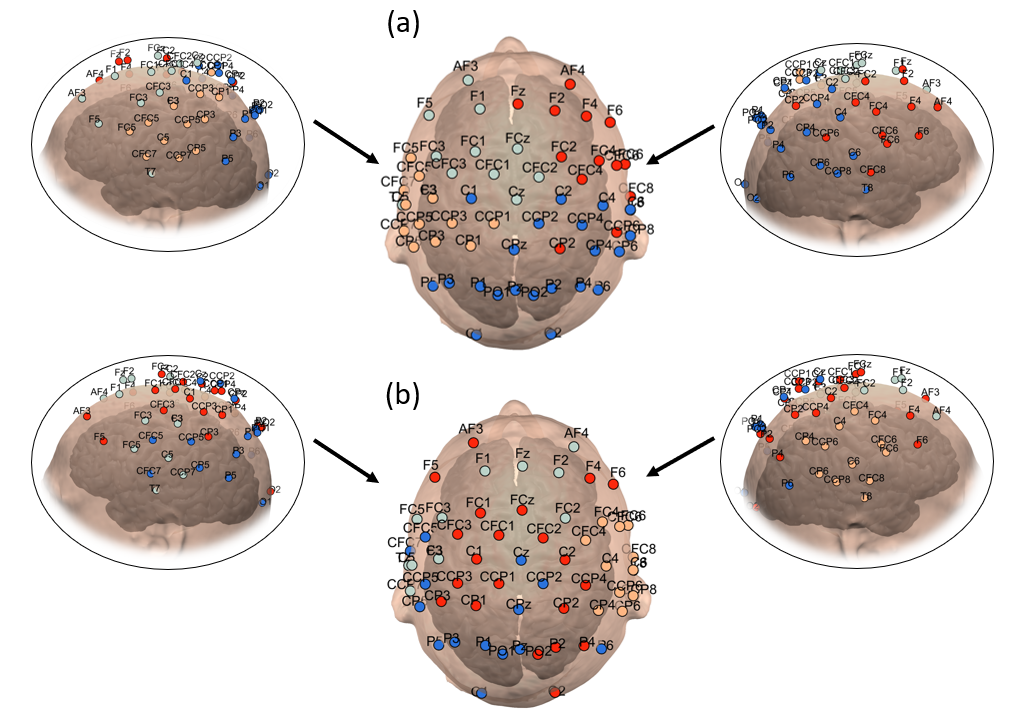}}
\caption{Output of community detection algorithm on real EEG data, where the colors of each node correspond to the resulting community. In the top line, we have results of EEG data recorded during motor imagery of the left hand, and in the bottom line we have those related to motor imagery of the right hand.
For this analysis, we previously selected $P = 4$ communities and we bandpassed signals in the alpha frequency band. }
    \label{fig:cd_real}
\end{figure*}

In order to investigate the organization of the communities on the brain scalp, we represent them in a different representation in the top line of Fig. \ref{fig:cd_real}.
In the three figures of panel (a), we have different views of the brain scalp where each node is coloured according to the associated communities. The identified clusters  appear symmetrically distributed in the brain and they geometrically follows the brain organization: we have a community in the frontal area, represented in blue, a community localized in the central electrodes, in light blue, then we have two symmetric communities in the two hemispheres, i.e. coloured in orange and in red, and a community that contains nodes in parietal and temporal electrodes (represented in yellow).

We now compare the cluster found during resting state with another condition, i.e. motor imagery of the right hand.
Results of the analysis related to the most representative trial are reported in the panel (b) of Fig. \ref{fig:cd_real}.
The output of the algorithm differs from the resting state condition. Indeed, here we remark the presence of the community in the left hemisphere that widens including all the left hemisphere in the contralateral motor area, that is supposed to be involved in the MI task, until the center and the beginning of the right hemisphere. 
Then we have a community in the frontal electrodes, two communities in the parietal lobe and one community, represented in orange, located in the right sensori-motor cortex. 
Our findings show that during motor imagery i) the symmetry of the brain communities is disrupted in the sensori-motor area, and ii) the community associated to the contralateral motor area widens.

Our approach  capture the variability of the human brain reconfiguration during motor imagery experiments. For this reason, brain clusters could be used as integrating feature to control a real Brain-Computer Interface by adding original insight as compared to the classical power spectral density or functional connectivity.

\section{Conclusions}
This paper proposes an original model to represent multiple observations, based on a multilayer network that can be written as the Cartesian Product Graph between the ground truth graph, that is the real graph underlying the possibly noisy observations, and a path graph, that has as many nodes as the number of layers. The key contribution of this work in the introduction of a novel approach to identify community structures in networked data, based on the estimation of optimized eigenvectors of the multilayer graph for community detection purpose. Our method has proved to outperform state-of-the-art competitors in identifying community structures on synthetic data. In addition, we apply the proposed framework to real EEG data collected during motor-imagery based BCI experiments and we demonstrate the potential of the method to highlight functionally relevant communities on the human brain. 

Taken together, the proposed method appears promising for modeling the brain EEG network and detecting community structures in the human brain during cognitive tasks, while being applicable in any case where differently estimated versions of a graph adjacency matrix are available and it paves the way for further generalization of the observation model. 

\begin{appendices}

\section{Computational Complexity}
The computational complexity of our method depends on the estimation of the first $N_{CD}$ eigenvectors. It can be approximated to $\mathcal{O}(N_{edges}\cdot N_{CD})$, where $N_{Edges}$ is the total number of edges and $N_{CD}$ is the number of eigenvectors that have to be estimated. 
In our case, the total number of edges depends both on the total number of intralayer links and the interlayer links.

Specifically, the total number of edges is intralayer links $N_e \cdot M$, because each layer has $N_e$ edges and, according to the proposed Cartesian Product Graph Model, the total number of intelayer connections is $N \cdot (M-1)$, where $N$ is the number of nodes of each graph and $M$ is the number of layers. The computational cost C corresponds to $C=N_e\cdot N\cdot M(M-1)$. Let us underline here that the graph is sparse, $N_e<<N\cdot N$.

\section{Label harmonization}
\begin{figure}
	\centering{\includegraphics[scale=0.73]{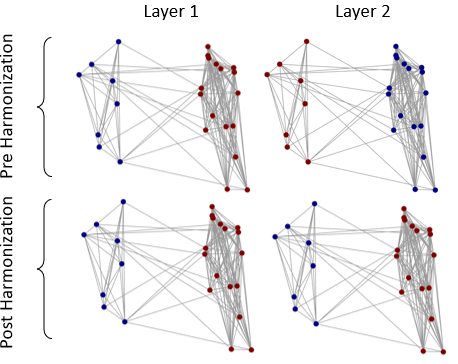}}
\caption{Example of harmonization result before majority voting. In this case, the parameters are: $N =25$, $P=2$, $M=2$, $N_{runs}=1$, $prob=0.9$. In the top line, we report the communities obtained with Louvain procedure with different colors before the harmonization. In the bottom line, we have the communities (identified with colors) after the harmonization step. }
    \label{fig:prepost}
\end{figure}

The centroids of the clusters found at the first iteration are computed and the subsequent  iterations the new centroids are compared with the stored ones and the name of the labels are possibly changed according to the smallest geometrical distance between novel and stored labels.
In Fig.\ref{fig:prepost}, we graphically represent the effect of the label harmonization. Indeed, in the top line, we report the clusters identified with the Louvain method, with  $N =25$, $P=2$, $M=2$, $N_{runs}=1$, $prob=0.9$ at the different layers. It is possible to remark that the clusters are exactly the same but the labels are shifted, since the community represented in blue at the first layer correspond to the red at the second layer. In the bottom line, we represent the cluster after the harmonization stage.

\end{appendices}

\section*{Acknowledgment}
FD acknowledges support from the European Research Council (ERC), the European Union’s through Horizon 2020 research and innovation programme under Grant $864729$

\bibliography{main}
\bibliographystyle{ieeetr}

\end{document}